\documentstyle[twoside,fleqn,espcrc2,psfig]{article}

\newcommand{\beq}{\begin{equation}}
\newcommand{\eeq}{\end{equation}}
\newcommand{\bea}{\begin{eqnarray}}
\newcommand{\eea}{\end{eqnarray}}

\newcommand{\chb}{\overline{\chi}}

\newcommand{\bt}{\beta}
\newcommand{\lag}{\langle}
\newcommand{\rag}{\rangle}
\newcommand{\gm}{\gamma}

\newcommand{\kp}{\kappa}
\newcommand{\lm}{\lambda}

\newcommand{\be}{\begin{equation}}
\newcommand{\ee}{\end{equation}}

\def \3{\ss}

\title{%
  \mbox{}\hbox to 0pt{\vbox to 0pt{\vss
    \parbox[b]{\hsize}{\normalsize
      \begin{flushright}
        HLRZ 56/95\\
        hep-lat/9509040
      \end{flushright}
      \vspace*{2.5cm}}}\hss}%
      Magnetic and chiral universality classes in a 3D Yukawa
      model\thanks{Work supported by the DFG and BMBF. Computations have
      been performed on the CRAY-YMP in J\"ulich and on the Quadrics QH2
      in Bielefeld.}}

\author{E. Focht\address{Institute for Theoretical Physics E,
            RWTH-Aachen, 52074 Aachen, Germany\\
            HLRZ c/o KFA J\"ulich, 52425 J\"ulich, Germany},
        J. Jers\'ak$^{\rm a}$
        and
        J. Paul$^{\rm a}$
}

\begin{document}

\begin{abstract}
The 3D Yukawa model with U(1) chiral symmetry is investigated in a
broad interval of parameters using the Binder method. Critical
exponents of the Wilson-Fisher (magnetic) and Gross-Neveu (chiral)
universality classes are measured. The model is dominated by the
chiral universality class. However at weak coupling we observe a
crossover between both classes, manifested by difficulties with the
Binder method which otherwise works well.
\end{abstract}

\maketitle


\section{Introduction}

The existence of nontrivial fixed points in 4D is not yet
ruled out in nonperturbative calculations. Several gauge theories
have, in addition to the Gaussian fixed points, also suspicious
critical points at strong coupling. One of the purposes of our
investigation of the 3D Yukawa model (Y$_3$) was to learn how to deal
with models having several nontrivial fixed points and crossovers
between the corresponding universality classes.

The Y$_3$ model is known to have two nontrivial fixed points: the
Wilson--Fisher fixed point of the pure scalar $\phi_3^4$ theory at
vanishing Yukawa coupling with a magnetic type phase transition, and
the fixed point of the 3D Gross--Neveu model (GN$_3$) with a chiral
phase transition.

The model is also interesting from the point of view of statistical
mechanics. Transitions between different universality classes have
been investigated in spin models \cite{DORe95} but not yet in fermionic
ones. The most promising method used is Binder's method of finite size
scaling analysis. We have applied this method succesfully to the
chiral phase transition. The failure of this method also indicates the
occurence of a crossover to the magnetic universality class.
The existence of intermediate universality
classes between the two, corresponding to the known fixed points, is
also of interest \cite{DORe95}, but in the Y$_3$ model we didn't detect
any signs for it.

\section{Lattice action and phase diagram}

We studied the Y$_3$ model on the lattice with staggered fermions,
hypercubic Yukawa coupling and U(1) chiral symmetry. The action is
\begin{eqnarray} \label{ACTION}
S &=& \sum_{x}
  \left[ -2 \kappa \sum\limits_{\mu}
    \phi_{x+ \mu}^{i} \phi_x^{i}
    + \phi_x^2 + \lambda ( \phi_x^2 - 1 )^2
  \right. \nonumber \\
&+& \frac{1}{2} \sum\limits_{\mu} \eta_{x,\mu}
  ( \chb_x^j \chi_{x+\mu}^j - \chb_{x+\mu}^j \chi_x^j ) \nonumber \\
&+& \left. y \chb_x^j \frac{1}{2^3}
 \sum\limits_b ( \phi_{x+b}^1 + i \varepsilon_x \phi_{x+b}^2 )
 \chi_x^j \right] \: .
\end{eqnarray}
The staggered sign factors are
$\eta_{x,1} = 1$, $\eta_{x,\mu} = (-1)^{x_1 + \ldots + x_{\mu - 1}}$
and $\varepsilon_x = (-1)^{x_1 + \ldots + x_d}$. The indices $x+\mu$
and $x+b$ denote, respectively, the nearest neighbors of the site $x$
and the corners of the associated elementary cube. The scalar field
$\phi_x$ is a two component real field.

\begin{figure}
  \begin{center}
    \leavevmode
    \psfig{file=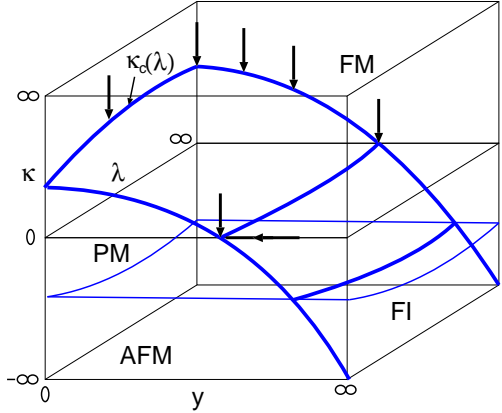,height=5.5cm}
  \end{center}
  \vspace{-1cm}
  \caption{ Sketch of the Y$_3$ phase diagram.}
  \label{FIG:PD}
\end{figure}

{}From considerations of the effective potential to 1-loop order and
numerical simulations we determined the phase diagram sketched in
fig.~\ref{FIG:PD}. It has two well-known limits: the $\phi_3^4$ model
at $y=0$ and the GN$_3$ model at $\lm=0$, $\kp=0$.
The region below the critical surface is the paramagnetic phase (PM),
the region above the ferromagnetic phase (FM). For negative values
of the parameter $\kp$ we further expect an antiferromagnetic phase
(AFM) and a ferrimagnetic phase (FI).

\section{Numerical methods}

For the numerical simulation we used a hybrid Monte-Carlo program.
Ferrenberg-Swendsen multihistogram reweighting was used to interpolate
between the measured points in the parameter space.

In order to find an appropriate method to identify the universality
class corresponding to a certain critical point we tried several
methods to compute critical exponents in the pure scalar limit of the
model ($y=0$). The direct method using the scaling laws of the
physical quantities and the method using the Lee--Yang zeros of the
partition function needed significantly more statistics compared to
Binder's method of finite size scaling analysis of cumulants. Thus we
used only the Binder method in the fermionic case, with the
following definition for the cumulant:
\begin{equation}
U_L=2-\frac{\sum\limits_{x_1,\dots,x_4}\sum\limits_{i,j}
\lag\phi_{x_1}^i\phi_{x_2}^i\phi_{x_3}^j\phi_{x_4}^j\rag}{
(\sum\limits_{x_1,x_2}\sum\limits_{i}\lag\phi_{x_1}^i\phi_{x_2}^i\rag)^2}
{}~.
\label{CUMULANT}
\end{equation}
The validity of the hyperscaling hypothesis implies that these
cumulants are independent of the lattice size at the critical
point. This is a very precise way of determining critical couplings.

Critical exponents can be computed by considering pairs $(bL,L)$ of
lattice sizes. The cumulants deliver the exponent $\nu$ of the scalar
correlation length $\xi$:
\begin{equation}
  \ln\frac{\partial U_{bL}}{\partial
    U_L}\bigg|_{\kappa_c} = \frac{1}{\nu} \ln b\; .
  \label{EQ:NU}
\end{equation}
Similarly we obtained $\gm/\nu$ and $\bt/\nu$ from the susceptibility
$\chi_L$ and the magnetization $M_L$.

\section{The $\phi_3^4$ model}

The action (\ref{ACTION}) describes at vanishing $y$ free massless
fermions and $O(2)$ invariant scalars with selfinteraction ($\phi_3^4$
model). We have determined the phase diagram of this model (for the
data see \cite{FoJe95}). At positive $\kp$ a critical line of second
order phase transitions $\kp_c(\lm)$ separates the PM and FM phases.
At the point $\kp_c(\lm=0)$ the theory is dominated by a Gaussian
fixed point, being asymptotically free at large momenta. All other
points on the critical line $\kp_c(\lm>0)$ lead to a nontrivial
continuum limit, dominated by the Wilson--Fisher fixed point. In order
to test this expectation, we measured the renormalized quartic coupling
and the critical exponents $\nu$, $\bt/\nu$ and $\gm/\nu$ at two
values of the bare scalar quartic coupling: $\lm=0.5$ and
$\lm=\infty$. Fig. 3a illustrates the quality of
determination of $\kp_c$ by the Binder method at $\lambda
= \infty$. These measurements helped us to develop and test the
methods later used for nonvanishing $y$.

The renormalized scalar quartic coupling
\begin{equation}
  \lm_R=(Lam_R)^3U_L
  \label{LAMBDA:R}
\end{equation}
has been determined by keeping the ratio $L/\xi$ fixed to $4$ and
varying the lattice size $L$ between $6$ and $12$. The extrapolation
to infinite volume suggests $\lm_R=26(4)$ for both values of
$\lm$.

The critical exponents we calculated by the Binder method are
summarized in the following table.
\begin{center}
  \leavevmode
  \begin{tabular}{|c|c|c|c|c|}
    \hline
    $\lambda$ & $\kappa_c$ & $\nu$ & $\beta/\nu$ & $\gamma/\nu$\\
    \hline
    $\infty$ & 0.2275(10) & 0.673(19) & 0.51(3) & 2.03(6) \\
    0.5 & 0.241(1) & 0.687(19) & 0.56(5) & 1.91(6) \\
    \hline
  \end{tabular}
\end{center}
The results are consistent within error bars and also with the
hyperscaling hypothesis. They confirm that the continuum limits at
$\lm=\infty$ and $\lm=0.5$ belong to the Wilson--Fisher universality
class.

\section{Y$_3$ model at $\lm=0$}

The GN$_3$ model with $U(1)$ chiral symmetry arises in the limit
$\lm=0$. It is renormalizable in the $1/N_F$ expansion and its $\bt$
function has been calculated up to $O(1/N_F^2)$ \cite{Gr94}. The
critical exponent $\nu$ resulting from these calculations is
$\nu\simeq 1.0(1)$.

The phase structure of the GN$_3$ model can be read of
fig. \ref{FIG:PD}. The symmetric phase ($y<y_c$), where fermions
are massless, is dominated by the Gaussian fixed point at $y=0$.
The critical point $y_c=1.091(5)$ is an UV stable nontrivial fixed
point.

\begin{figure}[htbp]
  \begin{center}
    \leavevmode
    \psfig{file=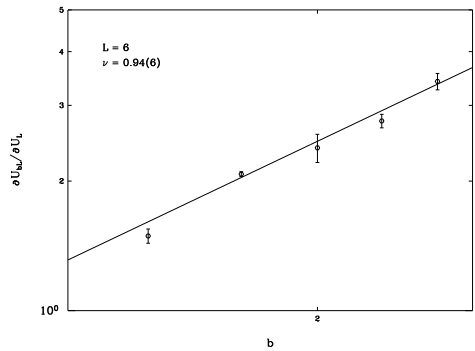,angle=90,height=5cm}
  \end{center}
  \vspace{-1cm}
  \caption{Determination of the critical exponent $\nu$ in the
    Gross--Neveu.}
  \vspace{-6mm}
  \label{FIG:GN:NU}
\end{figure}

We applied the Binder method to the GN$_3$ model. We approached the
critical point both by varying $y$ at $\lm=\kp=0$ (GN case) and $\kp$
at $\lm=0$ and $y=y_c$. Fig. \ref{FIG:GN:NU} shows the
determination of $\nu$ using eq. (\ref{EQ:NU}). Fig. 3b
shows that the cumulants cross in the GN case at $y=1.091$.
The obtained exponents are perfectly consistent with each other and
the theoretical values. Their averages are: $\nu=1.03(10)$,
$\bt/\nu=0.89(7)$, $\gm/\nu=1.17(9)$,
values which are significantly different from the Wilson--Fisher
exponents and allow the investigation of crossover effects between
these universality classes.

  \begin{center}
    \leavevmode
    \psfig{file=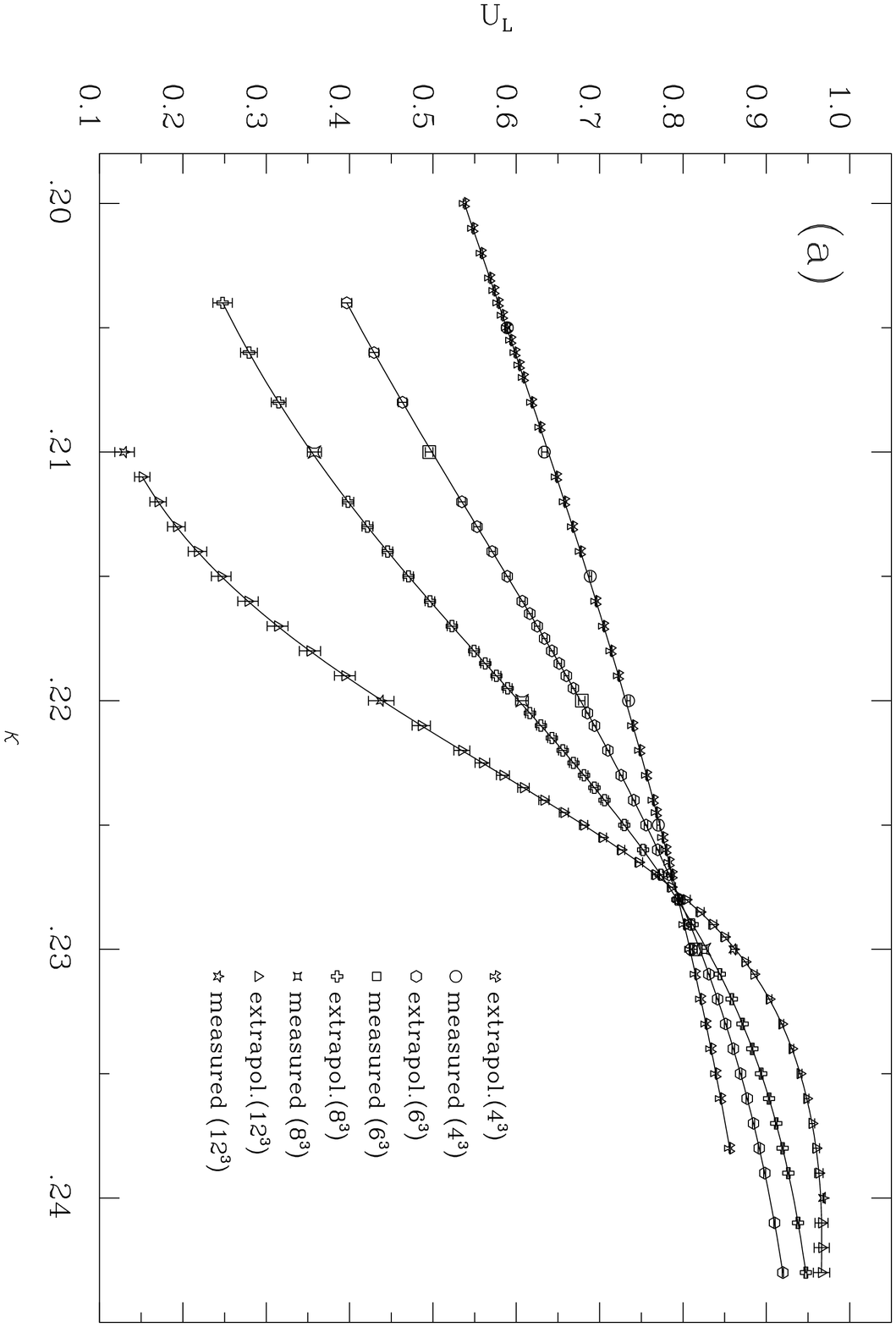,angle=90,height=4.5cm}
    \psfig{file=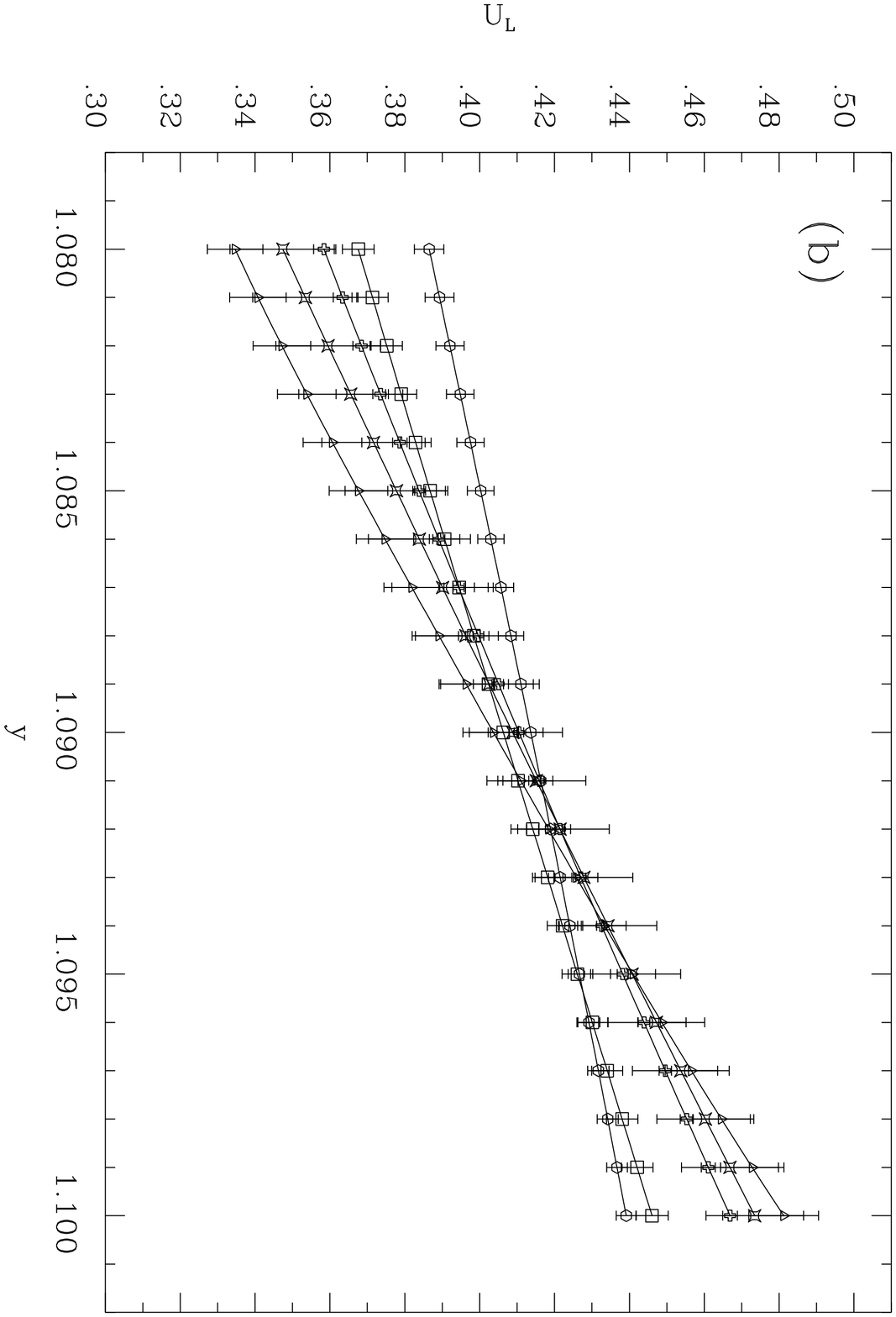,angle=90,height=4.5cm}
    \psfig{file=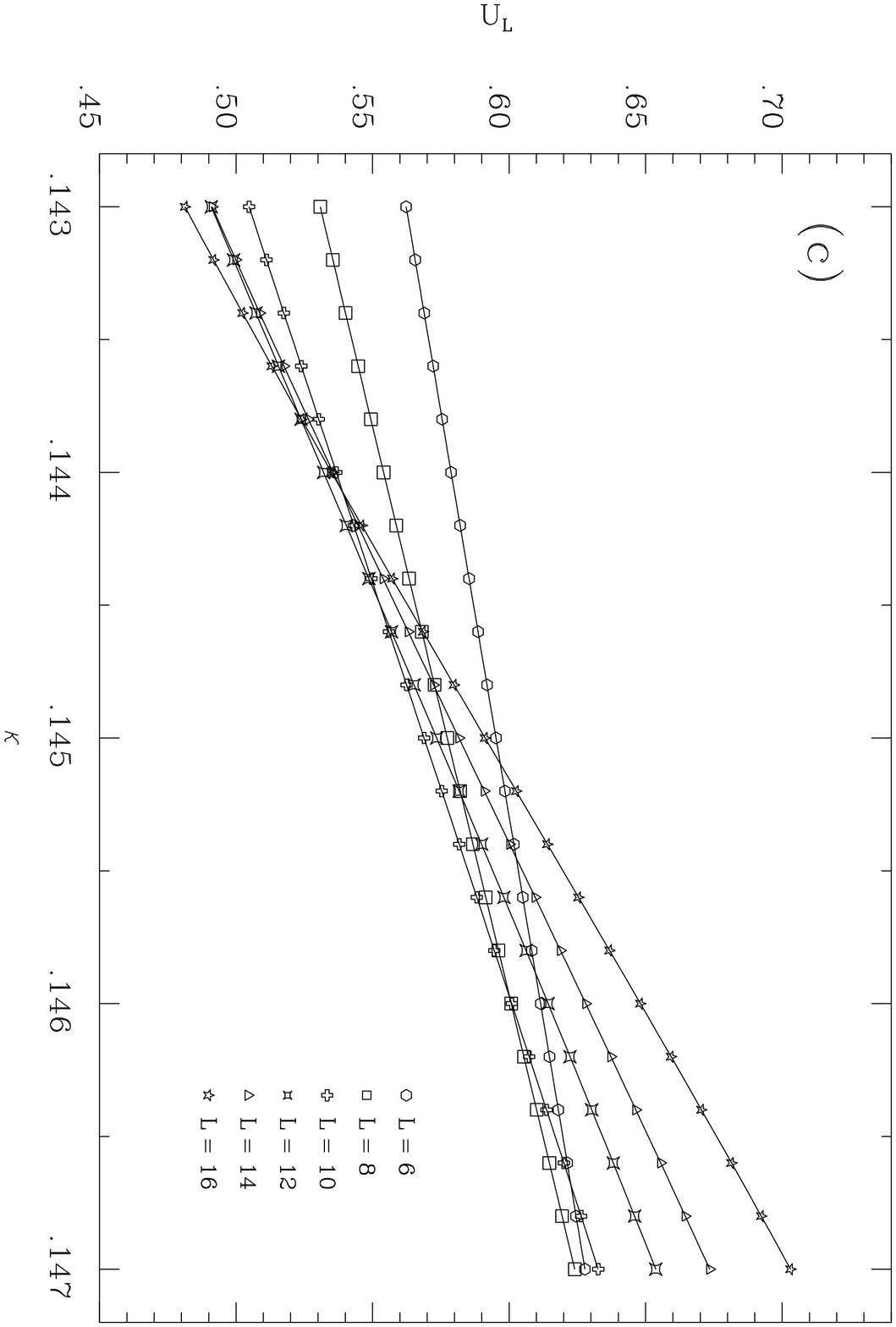,angle=90,height=4.5cm}
  \end{center}
  Figure 3. Determination of $\kp_c$: (a) in the XY$_3$ model,
    (b) in the GN$_3$ model and (c) at $\lm=\infty$, $y=0.6$. It works
    well in the first two cases, but only poorly in the third case.\par
\vskip 7mm

\addtocounter{figure}{1}

\section{Y$_3$ model at $\lm=\infty$}

There are theoretical arguments \cite{Zi91} which predict that Y$_3$
and GN$_3$ models belong to the same universality class for small
values of $\lm$. We tested this conjecture for $\lm=\infty$ by
determining the critical exponents at large Yukawa coupling
($y=1.1$).

The Binder method works as well as in the GN$_3$ case. At $y=1.1$ and
$\lm=\infty$ we varied $\kp$ and determined first its critical
value. The intersection point of the cumulants on different lattice
sizes delivers with good accuracy $\kp_c=0.007(2)$. Using
eq. (\ref{EQ:NU}) the critical exponent $\nu$ has been extracted:
$\nu=0.89(6)$. Further we determined: $\bt/\nu=0.80(8)$ and
$\gm/\nu=1.30(7)$.

All these exponents are consistent with those of the GN$_3$ fixed
point and significantly different from the Wilson--Fisher
exponents. We conclude that, at least for strong enough Yukawa
coupling, the critical surface from $\lm=0$ to $\lm=\infty$ belongs to
the GN universality class.


At $\lm=\infty$ but smaller Yukawa coupling crossover phenomena make
the determination of critical exponents more difficult.

At $y=0.6$ it is difficult to determine $\kp_c$
(fig. 3c). For small lattices
($L=6,8,10$) the cumulants cross in the interval
$\kp=0.146-0.1466$. The finite size analysis delivers $\nu=0.79(9)$,
inconsistently to the GN universality class. When only lattices larger
than $L=10$ are considered, the crossing point of the cumulants is
$\kp_c\simeq 0.144$. An analysis for this $\kp_c$ leads to
$\nu=0.99(23)$. This value is consistent with the GN critical exponent.
The reason for the strong dependence of the calculated $\nu$ on
$\kp_c$ is a larger curvature in the functions $U_{bL}(U_L)$. Thus the
value of the derivative $\partial U_{bL}/\partial U_L$ depends
stronger on the value of $\kp_c$ than in the $\phi_3^4$ and strong $y$
cases.

The computations at $y=0.3$ revealed even stronger crossover
effects. Though lattices up to $L=32$ have been used, the value of
$\kp_c$ couldn't be determined. A very strong curvature in
$U_{bL}(U_L)$ makes an accurate  calculation of $\nu$ impossible on
such small lattices.

\section{Boson mass in the Y$_3$ model}

In the Gross--Neveu model the bosonic particles can be interpreted as
fermion-antifermion bound states, because the scalar field introduced
in the action is auxilliary. In the full Yukawa model and its purely
scalar limit $\phi_3^4$ the $\phi$ field is dynamical. We measured the
propagator of the two components of the $\phi$ field in momentum
space. While in the $\phi_3^4$ model this propagator can be fitted
with a free boson Ansatz, its form is very complex at nonvanishing
Yukawa coupling. We were able to fit it with an Ansatz using
renormalized lattice perturbation theory.

\begin{figure}[htbp]
  \begin{center}
    \leavevmode
    \psfig{file=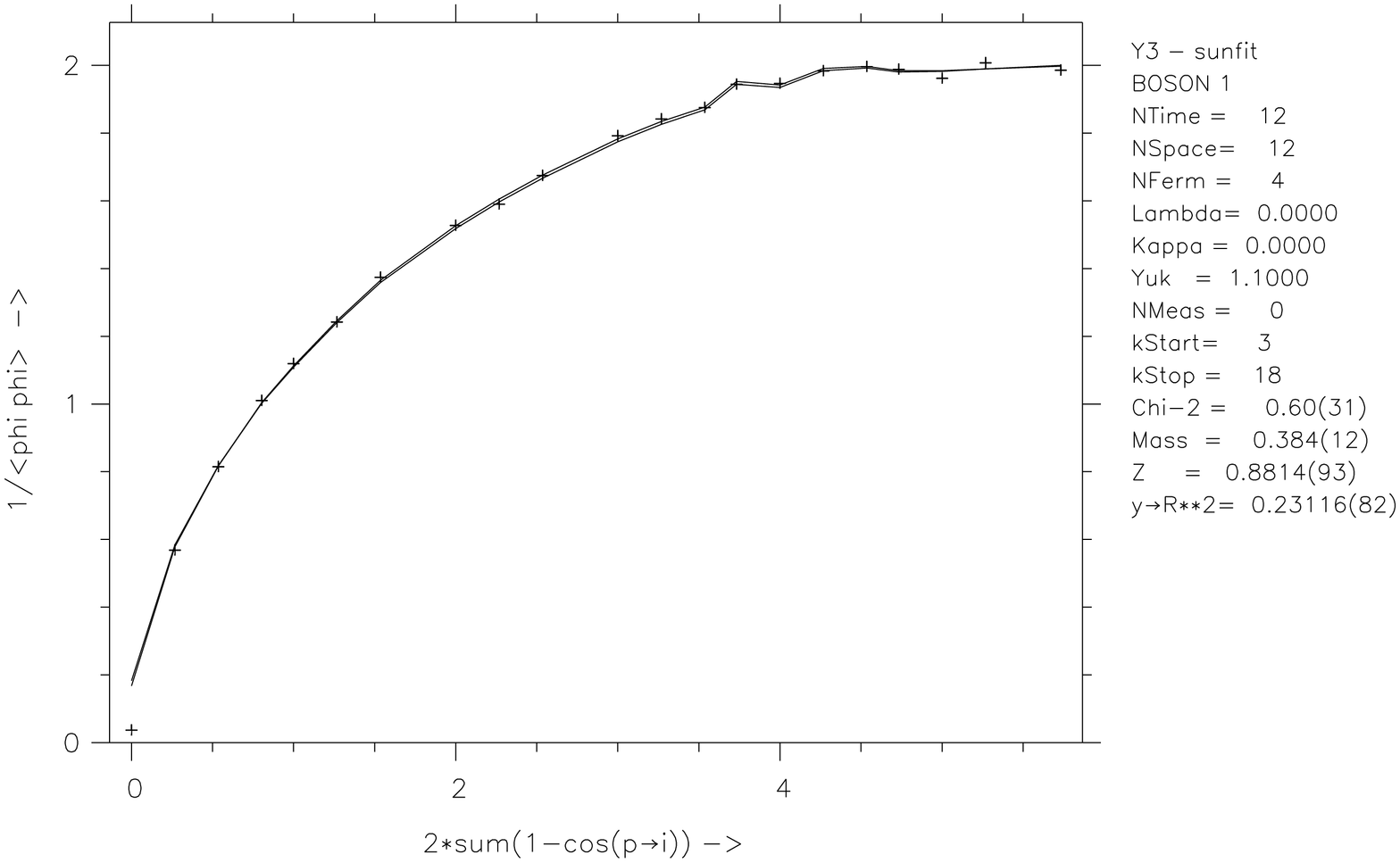,angle=90,height=5cm}
  \end{center}
  \vspace{-1cm}
  \caption{Bosonic propagator in the GN$_3$ model at
    $\lm=\kp=0$, $y=1.1$.}
  \vspace{-6mm}
  \label{FIG:SC}
\end{figure}

Fig. \ref{FIG:SC} shows such a bosonic propagator in the Gross--Neveu
case and the fit. The characteristic form, which is reproduced well by
the fit, is the same for all $\lm$ values, including
$\lm=\infty$. Also the behavior of the boson mass across the phase
transition is very similar both at $\lm=0$ and $\lm=\infty$.
This is another piece of evidence that the Y$_3$ and GN$_3$ models are
equivalent (belong to the same universality class).

\end{document}